# OPTIMAM Mammography Image Database: a large scale resource of mammography images and clinical data

Mark D Halling-Brown[1,2], Lucy M Warren[1], Dominic Ward[1], Emma Lewis[1, 2], Alistair Mackenzie[1], Matthew G Wallis[3], Louise Wilkinson[4], Rosalind M Given-Wilson[5], Rita McAvinchey[6], Kenneth C Young[1,2]

[1]Royal Surrey NHS Foundation Trust, Egerton Road, Guildford, GU2 7XX, UK
[2]University of Surrey, Guildford, GU2 7XH, UK
[3]Cambridge Breast Unit, Cambridge University Hospitals NHS Foundation Trust, Cambridge & NIHR Cambridge Biomedical Research Centre, Cambridge, CB2 0QQ, UK
[4]Oxford Breast Imaging Centre, Oxford University Hospitals NHS Foundation Trust, OX3 7LE, UK
[5]Department of Radiology, St George's Healthcare NHS Trust, London, SW17 0QT, UK
[6]Jarvis Breast Screening Unit, Guildford, GU1 1LJ, UK

**Corresponding author details**
Dr Mark Halling-Brown, Royal Surrey NHS Foundation Trust, Egerton Road, Guildford, GU2 7XX, UK +44(0)1483 571122 (Ext: 6876), mhalling-brown@nhs.net

## Abstract

A major barrier to medical imaging research and in particular the development of artificial intelligence (AI) is a lack of large databases of medical images which share images with other researchers [1].  Without such databases it is not possible to train generalisable AI algorithms, and large amounts of time and funding is spent collecting smaller datasets at individual research centres.  The OPTIMAM image database (OMI-DB) has been developed to overcome these barriers.  OMI-DB consists of several relational databases and cloud storage systems, containing mammography images and associated clinical and pathological information.  The database contains over 2.5 million images from 173,319 women collected from three UK breast screening centres.  This includes 154,832 women with normal breasts, 6909 women with benign findings, 9690 women with screen-detected cancers and 1888 women with interval cancers.  Collection is on-going and all women are followed-up and their clinical status updated according to subsequent screening episodes.  The availability of prior screening mammograms and interval cancers is a vital resource for AI development.  Data from OMI-DB has been shared with over 30 research groups and companies, since 2014.  This progressive approach has been possible through sharing agreements between the funder and approved academic and commercial research groups.   A research dataset such as the OMI-DB provides a powerful resource for research.

## Introduction

The development of AI software products to improve the outcomes of breast screening is reliant on the availability of well-curated image databases [1]. The OPTIMAM Mammography Image Database (OMI-DB) [2, 3] was created to provide a centralized, fully annotated dataset for research. The initial reason for creating the database was to conduct research in the Cancer Research UK funded projects OPTIMAM (2008-2013) and OPTIMAM2 (2013-2018) which evaluated how various factors affect breast cancer detection in mammograms. In the UK, the National Health Service Breast Screening Programme (NHSBSP) invites women to attend breast screening every three years between the ages of 50 and 70. At some screening centres younger and older women are also invited for screening as part of the national age trial [4]. Some women in high-risk groups may receive annual invitations to screening. Our objective was to collect all mammograms for women with screen detected cancers at three screening centres (sites) as well as representative samples of normal and benign screening cases. This resource was designed to be shared for research purposes, most notably in Virtual Clinical Trials (VCT), Computer Aided Detection (CAD), artificial intelligence (AI), image perception studies, training and quality assurance.

## Image database: Image collection and design

The processes and systems required to allow image collection are complex. Figure 1a shows a simplified view of the two types of image collection: *automated remote-site* and *stand-alone*. A full description of the processes can be found in the Supplementary Material. The collection site's clinical database, the National Breast Screening System (NBSS), is queried to identify a set of clients based on collection requirements, for example study-date range, outcome classification (Normal or Malignant), or high-risk status. Images and clinical data for the clients' screening and assessment episodes are retrieved. For all the images currently in OMI-DB from three screening centres, this process has been fully automated using a physical or virtual server on site. However, tools have been developed and tested for future collection sites where setting up a server is not possible or feasible. In this stand-alone collection, the image collection tool is downloaded by a staff-member and pointed at a manually prepared folder containing images for collection, and the NBSS databases queried for clinical data for those clients. All further processing and storage of imaging and NBSS data follows the automated collection procedure.

Imaging and screening data is pseudonymised and records inserted into the clinical collection site's lookup tables. Images, image meta-data and screening data are uploaded to the cloud for storage in a central database. All collection activity relating to clients, e.g. NBSS and PACS queries, and meta-data related to collection runs, is logged at the collection site.

The OMI-DB is made up of several relational databases and cloud storage systems [3]. Figure 1b shows an amalgamated, simplified schema of the data model. The associated data comprises radiological, clinical and pathological information extracted from the NBSS.

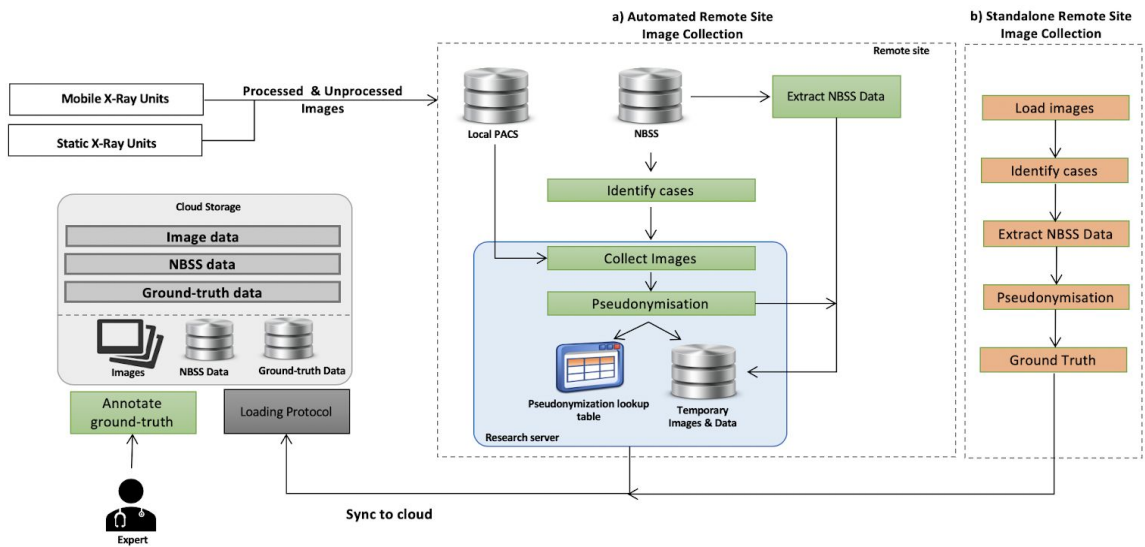

Figure 1a: simplified representation of processes for collecting and annotating imaging, clinical and ground truth data used to populate the OMI-DB

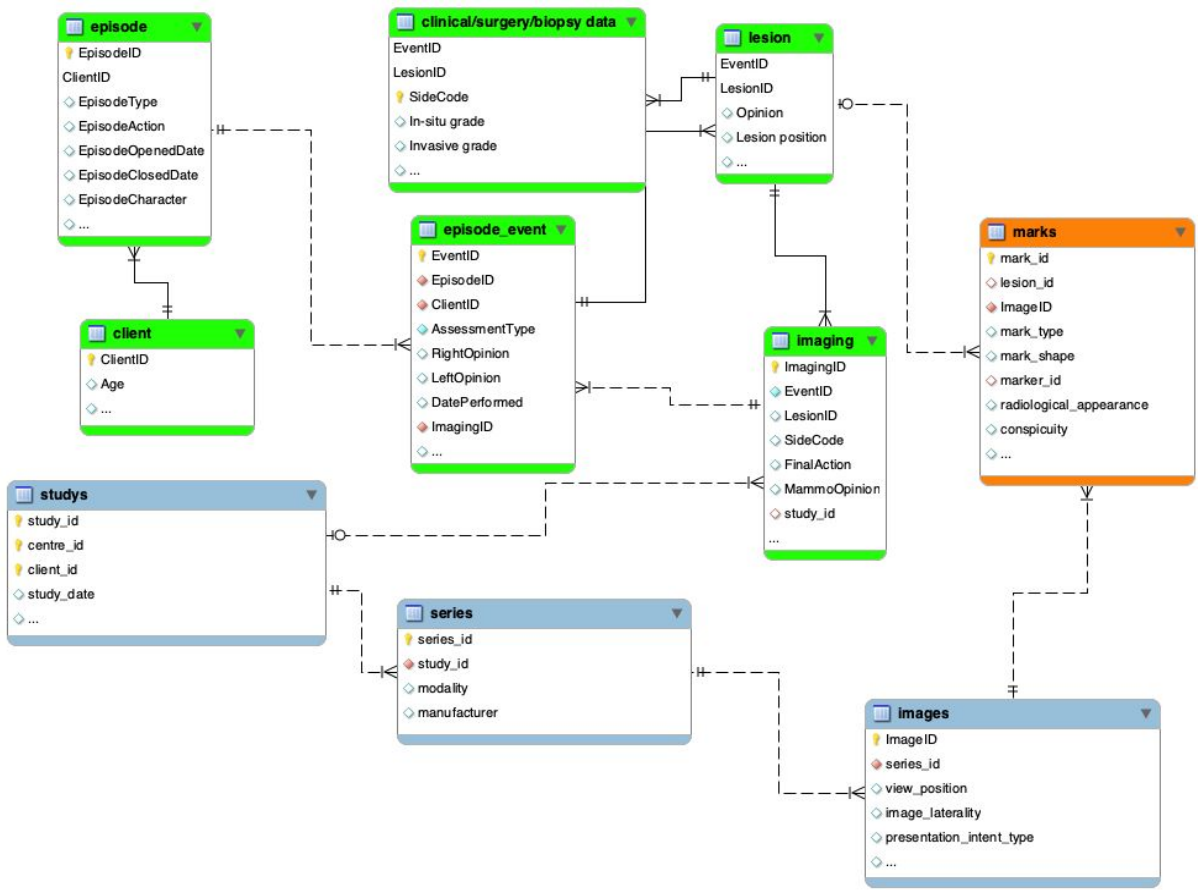

Figure 1b: schema showing simplified data model for radiological (tables marked in blue), clinical and pathological (tables marked in green) and ground-truth (tables marked in orange) information stored in the OMI-DB

When loading the images into the image database, all relevant DICOM tags are extracted to allow a searchable index to be produced. Information on screening history, previous occurrences of cancer, biopsy results and surgical procedures are collected from NBSS. The radiological location of lesions are not stored routinely in the clinical databases. However, such information is particularly useful for training and evaluating AI algorithms. This has been collected specifically for OMI-DB. Experienced mammography readers have drawn regions-of-interest indicating the location and area of lesions and other attributes, such as radiological appearance and conspicuity. 7143 lesions have been marked (including approximately 60% of screen-detected cancers). A web-enabled, remotely accessible software application has been developed which allows radiologists to view cases, annotate clinical features and participate in observer studies [5].

## Content of OMI-DB

OMI-DB contains images and associated clinical data that has been collected since 2011 from three UK screening sites - Jarvis Breast Screening Centre in Guildford, St George's Hospital in South West London and Addenbrookes Hospital in Cambridge.

Screening mammograms for all screen-detected cancers and the prior screening mammograms of interval cancers since 2012 have been collected from the three screening sites. In addition, images and clinical data were collected for all women screened between 1st January 2014 and 31st December 2014, and 25% of all women screened in 2012, 2013 and 2015 were randomly selected for database inclusion. The total number of all types of images in the database is 2,889,312.

The database contains unprocessed and processed images. Unprocessed images are essential for studies on image processing, and how the design parameters for imaging systems affect clinical performance and CAD.

Table 1 provides the distribution of invasive status and grade for screen-detected cancers and interval cancers, calculated from the NBSS clinical annotations. Twenty screen-detected cancers and twelve interval cancers were excluded due to lack of information on invasive status and grade in the clinical databases.

The availability of previous screening events and interval cancers opens up a wealth of potential research applications evaluating whether an abnormality could have been detected on the previous screening images via alterations to processing or perception. The number of women with 1, 2, 3 or more screening episodes with images in OMI-DB are shown in Figure 2. In addition to these episodes with images, OMI-DB contains clinical information for all episodes prior to when imaging systems at the clinical sites started using digital systems.

Initially, collection was only for 2D digital mammography images, however as the collection has progressed we have included additional modalities: such as tomosynthesis and MRI.

Table 1. Distribution of invasive status and grade, of screen-detected and interval cancers in OMI-DB[1]

| Invasive status | Grade | Screen detected cancers | Interval cancers |
|---|---|---|---|
| Invasive | Grade 1 | 1593 | 148 |
| | Grade 2 | 3518 | 610 |
| | Grade 3 | 1212 | 408 |
| | Not Assessable | 39 | 6 |
| | No grade | 556 | 178 |
| In-situ | Low grade | 205 | 13 |
| | Intermediate grade | 673 | 36 |
| | High grade | 1205 | 56 |
| | Not Assessable | 0 | 0 |
| | No grade | 515 | 36 |
| **TOTAL** | | 9516 | 1491 |

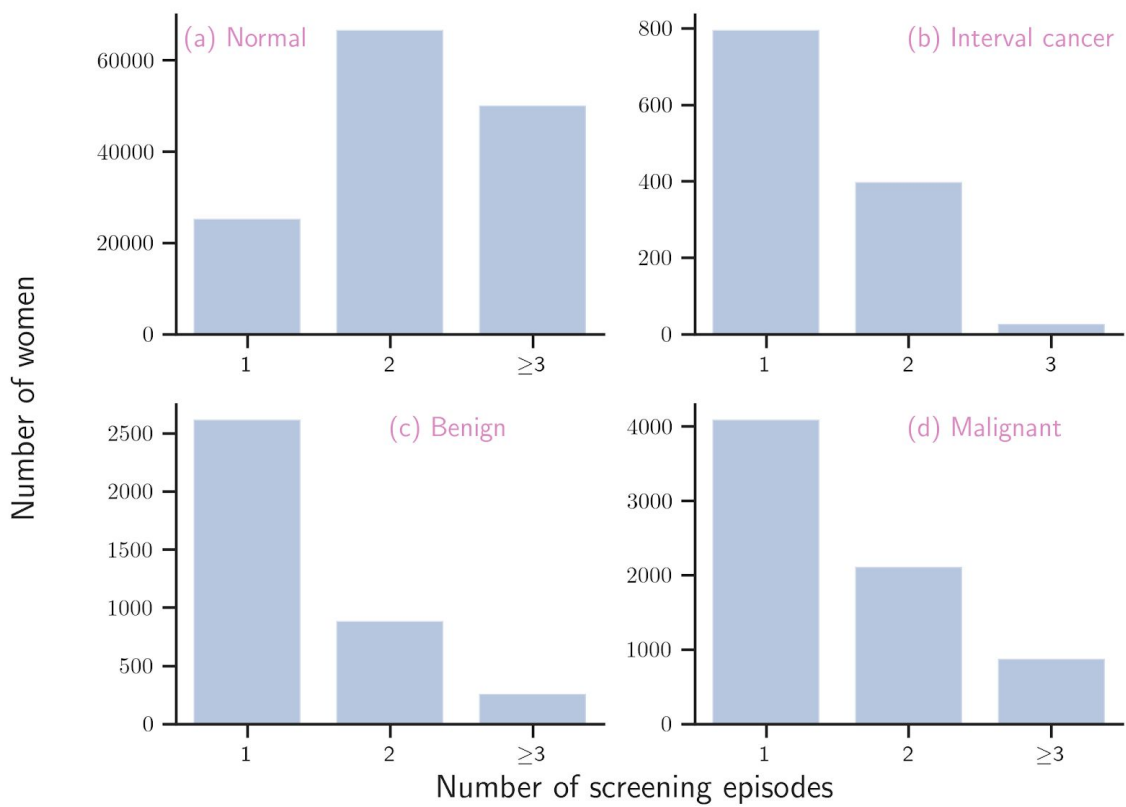

Figure 2. Number of women with 1, 2, 3 or more screening episodes with images in OMI-DB (a) Women with normal breasts (b) Women with interval cancers (c) Women with benign lesions (d) Women with screen-detected cancers

---

[1] For details of the classification procedures and source code used to generate the count data, visit https://bitbucket.org/scicomcore/omidb-2020

## Data Sharing

The project has approval from an ethical research committee specializing in research databases organised by the NHS Health Research Agency. The funder (Cancer Research UK) retains the intellectual property of the database and implements sharing agreements with approved academic and commercial research groups. This is a progressive approach to data collection and centralization and benefits the wider scientific community by avoiding the need for each group to undertake the enormous task of collecting their own datasets. The sharing of images and metadata with external parties requires additional processes involving further de-identification. A dedicated database records cases that have been shared with each third-party. The database also keeps a record of the primary investigators for the third party and additional information pertinent to sharing. A dedicated download co-ordinating tool has been developed to facilitate secure transfer and regularly synchronizes the metadata defined by the access list. Researchers are encouraged to use our open-source Python package that parses the shared OMI-DB to provide a detailed API and tools to facilitate metadata extraction and filtering. For enquiries regarding accessing data and images from OMI-DB, contact the CRUK Commercial Partnerships Team or apply via the OMI-DB website [14].

## Use of database

The database has been used in many projects undertaken at the Royal Surrey NHS Foundation Trust [6-9]. This includes virtual clinical trials to investigate the effect of factors such as detector type, dose and image processing on breast cancer detection [8,9]. Finally, images and data from OMI-DB have been used to evaluate the cancer characteristics and breast density of women in the UK NHSBSP [7, 10].

The OMI-DB has been shared widely to many groups for a variety of research aims. The majority of the research has been to develop machine-learning techniques applied to mammography images. Selected data from the OMI-DB has been shared with over 30 academic, research and commercial groups to train their AI algorithms.

As well as for training AI algorithms, it has been possible to share independent images for evaluation of AI algorithms that have not been used for other purposes. Images and data have been used to evaluate several AI algorithms at different stages of development from prototypes to CE marked products available for purchase [11-13].

## Discussion

The difficulty and cost of setting up an annotated mammographic image database with sharing protocols should not be underestimated. Any collection process should ideally be automatic, link clinical data to the images, whilst retaining confidentiality and expert annotation. Developing such a system has been time consuming and challenging. Since 2011, we have met these challenges and created a large database of images and associated data of the full range of cases acquired during breast screening. Collection into the database is ongoing and it is updated with any new information or subsequent screening episode for each case. Live updates on the data presented in Table 1 and Figure 2 are provided at the OPTIMAM database website [14]. The database has sharing protocols that allow the images to be used by researchers around the world [11-13].

The availability of previous screening events and interval cancers opens up a wealth of potential AI research applications evaluating whether an abnormality could have been detected on the previous screening images. A research dataset such as the OMI-DB provides a powerful resource for research. The sequential normal cases can be analysed using quantitative imaging features, with a priori knowledge that some years later these cases develop a malignancy.

Overall, a valuable, sharable database has been developed which holds both processed and unprocessed mammography images with annotated cancers and clinical details.

## Funding Information

The creation and development of the OPTIMAM image database is funded by Cancer Research UK (C30682/A28396).

## Acknowledgements

The authors would like to thank the Jarvis Breast Screening Centre, Rose Screen Centre at St George's Hospital and Cambridge Breast Unit at Addenbooke's Hospital for allowing copies of their images to contribute to OMI-DB. The authors would also like to thank Dr Vicky Cooke, Dr Julie Cooke, Dr Anna Cummin, Ms Charul Patel, Ms Pat Kelly, Ms Jacquiline Kenny and Ms Judy Leo for annotating the lesions within OMI-DB.

## Supplemental Materials

Website: https://medphys.royalsurrey.nhs.uk/omidb/
An open source Python API and command-line tool has been developed to facilitate and promote scientific research with OMI-DB. Project documentation can be found at https://scicomcore.bitbucket.io/omidb